\documentclass[12pt,a4paper]{article}
\usepackage{a4wide}
\usepackage{latexsym}
\usepackage{epsf}
\usepackage{amssymb}
\begin{document}
\def\be{\begin{equation}}
\def\ee{\end{equation}}
\def\bea{\begin{eqnarray}}
\def\eea{\end{eqnarray}}

\def\pd{\partial}
\def\a{\alpha}
\def\b{\beta}
\def\g{\gamma}
\def\d{\delta}
\def\m{\mu}
\def\n{\nu}
\def\t{\tau} 
\def\l{\lambda}
\def\s{\sigma}
\def\e{\epsilon}
\def\scri{\mathcal{J}}
\def\cM{\mathcal{M}}
\def\tcM{\tilde{\mathcal{M}}}
\def\RR{\mathbb{R}}
\def\CC{\mathbb{C}}

\hyphenation{re-pa-ra-me-tri-za-tion}
\hyphenation{trans-for-ma-tions}


\begin{flushright}
IFT-UAM/CSIC-00-32\\
hep-th/0011219\\
\end{flushright}

\vspace{1cm}

\begin{center}

{\bf\Large  Static Gauge Potential from Noncritical Strings}

\vspace{.5cm}

{\bf Enrique \'Alvarez}
\footnote{E-mail: {\tt enrique.alvarez@uam.es}}
{\bf and Juan Jos\'e Manjar\'{\i}n}
\footnote{E-mail: {\tt juanjose.manjarin@uam.es}} \\
\vspace{.3cm}

\vskip 0.4cm
 {\it   
Instituto de F\'{\i}sica Te\'orica, C-XVI,
  Universidad Aut\'onoma de Madrid \\
  E-28049-Madrid, Spain}\footnote{Unidad de Investigaci\'on Asociada
  al Centro de F\'{\i}sica Miguel Catal\'an (C.S.I.C.)}

\vskip 0.2cm

and
\vskip 0.4cm
{\it Departamento de F\'{\i}sica Te\'orica, C-XI,
  Universidad Aut\'onoma de Madrid \\
  E-28049-Madrid, Spain }

\vskip 0.2cm

\vskip 1cm


{\bf Abstract}
\end{center}
The static gauge potential between heavy sources is computed from
a recently introduced non-critical bosonic string background. When
the sources are located at the infinity of the holographic coordinate,
the linear dilaton behavior is recovered, which means that the potential
is exactly linear in the separation between the sources. When the sources are
moved towards the origin, a competing overconfining cubic branch appears,
which is however disfavored energetically.


\begin{quote}

\end{quote}


\newpage

\setcounter{page}{1}
\setcounter{footnote}{1}

\section{Introduction}
There has been recently, starting with the basic insight of Maldacena 
\cite{maldacena} several works purporting to the computation of Wilson loops
in different gauge theories.
\par
The general setup is as follows \cite{wilson}: there is a Poincar\'e invariant background,
of the type:
\be
ds^2 = a(r) d\vec{x}^2 + dr^2
\ee
where $ d\vec{x}^2$ stands for the flat metric in $d$-dimensional
Minkowski space (although we shall mostly work in the Euclidean version), 
$M_d$, and $r$ is the holographic coordinate.
\par
Then an idealized rectangular static loop is posited in spacetime,
by placing heavy sources at $x^1 \equiv x =\pm L/2$, $x^i = 0$ ($i\neq 1$).
\par
The basic postulate on which the computation relies \cite{cobi} consists
in representing it semiclassically through the minimal area surface
spanned by the world sheet of the confining string, assumed to have structure
in the holographic dimension only.
\par
To be specific, we identify world sheet and spacetime coordinates by choosing
\bea
x &&= \sigma\nonumber\\
t &&= \tau
\eea
and compute the minimal area through the Nambu-Goto action
\be
S_{NG}\equiv \int d^2 \sigma \sqrt{h}
\ee
where the induced metric, $h_{ab}$ is defined through
\be
h_{ab}\equiv \pd_a x^A  \pd_b x^B g_{AB}
\ee
(where we have represented world-sheet coordinates by $\sigma^a\equiv 
(\tau ,\sigma )$ and spacetime coordinates by $x^A$).
\par
The imbedding of the two-dimensional surface into the external spacetime
is defined by the unique function
\be
r = \bar{r}(\sigma )
\ee
determined through Euler-Lagrange's equations for the
Nambu-Goto action, complemented with Dirichlet boundary conditions
at the endpoints of the string,
\be
\bar{r}(0)=\bar{r}(1)=\Lambda
\ee
The resulting embedding then resembles half a topological cylinder,
with its length along the temporal direction, and bounded by $r=\Lambda$
(cf. Figure), in such a way that the total action is proportional to the span
of the temporal variable, $S\sim T$.
\begin{figure}[h]
\begin{center}
\leavevmode
\epsfxsize=8cm
\epsffile{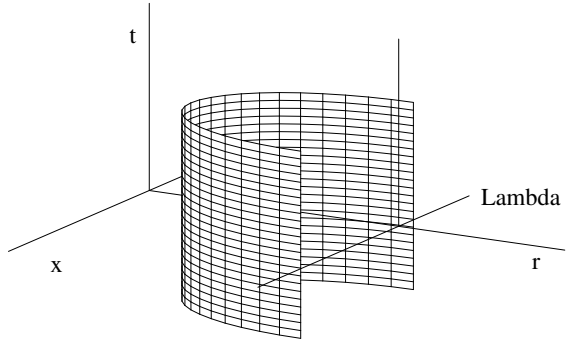}
\caption{\it Plot of the configuration}
\label{f4}
\end{center}
\end{figure}

\par

Finally, the potential between the heavy sources (a physical observable) is defined as
\be
V(L)\equiv lim_{T\rightarrow\infty}\frac{1}{T}S_{NG}
\ee
where $T$ is the time interval, and $S_{NG}$ is the on-shell value of the
Nambu-Goto action.
\par
In the original Ads/CFT correspondence, there are good reasons to choose
$\Lambda=\infty$ :there is a definite sense in which the CFT is defined
on an extension of Penrose's boundary, cf. {\cite{witten}, and, besides,
the mass of the sources is related to their
 location in the holographic direction.
\par
In other backgrounds this is not so clear, and one of the two main
 purposes of the 
present work is to study the dependence of the physical potential on 
the  value of $\Lambda$ (the location of the heavy sources in the holographic
direction).
\par
On the other hand, even in the first $AdS_5\times S_5$ background there are
extra dimensions which unavoidably lead to undesirable excitations. It is
clearly preferable in this context to have a background with the physical 
number of dimensions only.
Recently \cite{ag}, a new bosonic background has been introduced, which
makes sense in particular, in $d=4$ dimensions 
(plus one holographic coordinate). Its specific form is of the Poincar\'e
invariant type,
\be\label{metric}
ds^2= a(r)d\vec{x}^2 + dr^2 
\ee
with warp factor given by:
\be\label{warp}
a(r) = th(\sqrt{\frac{21}{6\a^{\prime}}}r)
\ee
and a dilaton background as well,
\be\label{dil}
\Phi = \frac{1}{2} \log \left[th(\sqrt{\frac{21}{6\a^{\prime}} }r) \frac{1}
{cosh^2 (\sqrt{\frac{21}{6\a^{\prime}} }r)}\right]
\ee

This background  interpolates between the well-known linear 
dilaton solution when
$r\rightarrow\infty$ and the confining background of \cite{ag} when $r\sim 0$.
Linear dilaton (which is exact to all orders in $\a'$) is just Minkowski
space,
\be
ds^2 = d\vec{x}^2 + dr^2
\ee
with $a=1$, and a dilaton field
\be
\Phi = \sqrt{\frac{21}{6\a'}}\, r
\ee
The confining background, on the other hand, is defined by:
\be
ds^2 = r d\vec{x}^2 + l_c^2 dr^2
\ee
(where $l_c$ is a length scale) and a dilaton given by
\be
\Phi = \log\, r
\ee
Were the matching between the two performed at a fixed value $r = r_0$,
the length scale $l_c$ would be fixed to be
\be
l_c\equiv \frac{1}{r_0}\sqrt{\frac{6\a'}{21}} 
\ee
The second main purpose of the present work is precisely
to study Wilson loops in this non-critical background.
\par
What we will find is that when the scale $\Lambda$ is big enough, the
effects of the confining background are not felt, which means that the
potential is purely linear (\cite{orlando}). As we
lower $\Lambda$ we reach a critical value below which the effects of the
confining background start to appear, which implies for the potential the
appearance of a competing, overconfining, cubic branch, which is not 
energetically favored.

\section{Wilson Loops in the Conformal Gauge}
The first thing we will  study is how to work in
the usual conformal gauge in Polyakov's action, 
\be 
S_{P}\equiv
\int \gamma_{ab}\pd^a x_{A}\pd^b x_{B} g^{AB} \sqrt{\gamma}d^2\sigma
\ee 
namely
$\gamma_{ab}\sim \delta_{ab}$.
\par

One can read the purported equivalence between Nambu-Goto and Polyakov
actions through the equations of motion. In the former case, avoiding
the factors $1/4\pi\alpha'$ in the Poyakov's case and $1/2\pi\alpha'$
in Nambu-Goto, these read

\begin{equation}
  \label{em1}
  \frac{\sqrt{h}}{2}h^{ab}\left[ 2g_{\mu\nu}(X^\alpha)\partial_a\partial_bX^\mu+2\partial_aX^\mu\partial_bX^\rho\partial_\rho g_{\mu\nu}(X^\alpha)+\partial_aX^\mu\partial_bX^\eta\partial_\nu g_{\mu\eta}(X^\alpha)\right]=0
\end{equation}

\noindent while for the Polyakov action one gets

\begin{equation}
  \label{em2}
   2g_{\mu\nu}(X^\alpha)\partial_a\partial^a X^\mu+2\partial_aX^\mu\partial^aX^\rho\partial_\rho g_{\mu\nu}(X^\alpha)+\partial_aX^\mu\partial^aX^\eta\partial_\nu g_{\mu\eta}(X^\alpha)=0
\end{equation}

\noindent It is plain that the equations (\ref{em1}) and (\ref{em2}) are equal if  the induced metric on the world-sheet is proportional to Kronecker's delta

\begin{equation}
  \label{prop}
  \frac{\sqrt{h}}{2}h^{ab}=\delta^{ab}
\end{equation}

The condition that the induced metric is proportional to the standard
 flat two dimensional Euclidean metric defines the so called 
isothermal coordinate system. That such a coordinate system can be 
locally defined is a standard result whose proof can be found, for example 
in \cite{spivak}. 

Precisely in this frame the non-linear sigma model action turns out to be 
the Dirichlet functional, and its stationary solutions can be interpreted 
as the surfaces of minimal area. This means that only in the isothermal 
coordinate system we can expect on general grounds an area law behavior when computing 
the Wilson loops.
\par
We can check explicitely this result in the case of AdS$_5\times$S$^5$, where the metric is given by

\begin{equation}
ds^2={\alpha}'\left( {\frac{U^2}{R^2}}dx^2+R^2{\frac{dU^2}{U^2}}+R^2d{\Omega}_{5}^2\right)
\end{equation}

\noindent and the actions are (neglecting the contribution of the $S_5$)

\begin{equation}
  \label{sp}
  S_P=\frac{1}{4\pi}\int_Md^2\sigma\left[ \frac{U^2}{R^2}(x')^2+\frac{U^2}{R^2}(\dot t)^2+ \frac{R^2}{U^2}(U')^2\right]
\end{equation}

\begin{equation}
  \label{ng}
  S_{NG}=\frac{1}{2\pi}\int_Md^2\sigma(\dot t)\sqrt{(U')^2+\frac{U^4}{R^4}(x')^2}
\end{equation}

\noindent The corresponding equations of motion are given by

\begin{eqnarray}
\partial_\sigma\partial^\s x+\frac{2}{U}\partial_\s U\partial_\s x=0\\
U^2\partial_\t\partial^\t t=0\\
\partial_\s\partial^\s U-\frac{1}{U}\partial_\s U\partial_\s U-\frac{U^3}{R^4}\left(\partial_\s x\partial_\s x+\partial_\t t\partial_\t t\right)=0
\end{eqnarray}

\noindent in the Polyakov's case, and by

\begin{eqnarray}
\partial_\s\left(\frac{U^4x'}{R^4\sqrt{(U')^2+\frac{U^4}{R^4}(x')^2}}\right)=0\\
\partial_\t\sqrt{(U')^2+\frac{U^4}{R^4}(x')^2}=0\\
\partial_\s\left(\frac{U'}{R^4\sqrt{(U')^2+\frac{U^4}{R^4}(x')^2}}\right)=\frac{2U^3x'^2}{R^4\sqrt{(U')^2+\frac{U^4}{R^4}(x')^2}}
\end{eqnarray}

\noindent in the Nambu-Goto one.

That this two results are identical is a simple consequence of (\ref{prop}).
\par

However, there is an interesting novel thing concerning the r\^ole played 
by the parameter $\s$ over the world-sheet. When working with the Nambu-Goto
 action, one was free to choose the static gauge, $\t=t$ and $\s=x$ 
(\cite{wilson}, \cite{ec1},\cite{cobi}). 
\par
Now we can no longer make this choice, 
because all gauge freedom has been used. What happens in the conformal
gauge is that the range of the world-sheet coordinate is dinamically
determined by the field equations:

\begin{equation}
\label{adssig}
\mathfrak{R}(\sigma)=\frac{R^2}{TU_0}\int_1^\infty\frac{dy}{\sqrt{y^4-1}}=\frac{R^2}{\sqrt{2}TU_0}F\left( \frac{\pi}{2},\frac{1}{\sqrt{2}}\right)
\end{equation}

It is useful for later purposes to represent it in terms of the minimal value
of the holographic coordinate, $U_0$, ( in case the boundary conditions
are located at  $U=\Lambda$,) that is:
\begin{equation} 
\mathfrak{R}(\sigma)=\frac{R^2}{\sqrt{2}TU_0}F\left( \cos^{-1}\left(\frac{U_0}{\Lambda}\right),\frac{1}{\sqrt{2}}\right) 
\end{equation}

\begin{figure}[h]
\begin{center}
\leavevmode
\epsfxsize=7cm
\epsffile{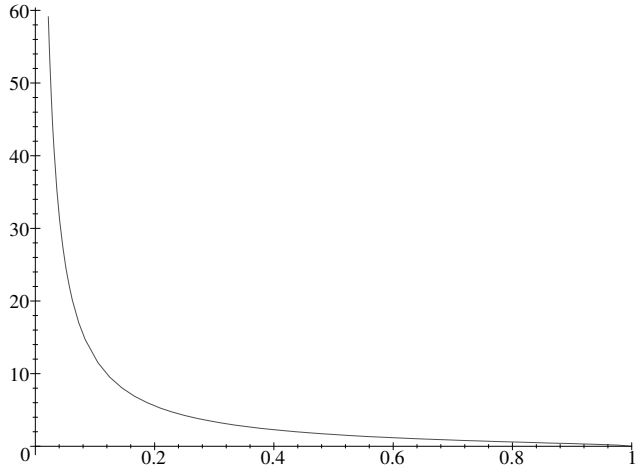}
\caption{\it $\frac{\mathfrak{R}(\sigma)T\Lambda}{R^2}$ vs. $\sqrt{\frac{U_0}{\Lambda}}$}
\label{f5}
\end{center}  
\end{figure}

\section{Wilson Loops in the Four-Dimensional Background}

Let us now turn to the novel  four-dimensional background recently obtained in
 \cite{ec1}. The metric reads

\begin{equation}
\label{metrlo}
ds^2=a(r)\eta_{\mu\nu}dx^\m dx^\n+dr^2
\end{equation}

\noindent where

\be
\label{ag1}
a(r)= \left[\frac{\sqrt{-c_1c_2}}{c_1}\tanh\left(
\sqrt{-c_1c_2}(r+c_3)\right)\right]^{2/\sqrt{d}}
\ee

\noindent for $c_1$, $c_2$ different from $0$, which is the case of interest
in the prersent context.
\par
In this background, the induced metric over the world sheet is

\be
ds^2_{\mbox{ind}}=\left[ a(r)(x')^2+(r')^2\right]d\s^2+a(r)(\dot t)^2d\t^2
\ee

The Polyakov action is

\be
S_p=\frac{1}{4\pi\alpha'}\int_M d^2\sigma\left[ a(r)\left( (x')^2+(\dot t)^2\right)+(r')^2\right]
\ee

The energy-momentum tensor takes the form

\be
j=a(r)\left( (x')^2-(\dot t)^2\right)+(r')^2
\ee

\noindent and the momentum 

\be 
p=a(r)x' 
\ee

\noindent they follow the relation $j=\frac{p^2}{a(r_0)}-a(r_0)(\dot t)^2$. We can define $j=\hat ja(r_0)(\dot t)^2$ and $p^2=\hat p^2a(r_0)^2(\dot t)^2$. Conformal invariance implies $\hat j=0$ and $\hat p^2=1$.

Using now the specific value for the warp factor (\ref{ag1}) one gets:

\be
\s=\frac{c_1^{1/\sqrt{d}}}{(\dot t)\left( -c_1c_2\right)^{(1+\sqrt{d})/(2\sqrt{d})}\tanh^{(1-\sqrt{d})/\sqrt{d}}(v_0)}\int_1^Y\frac{y^{(\sqrt{d}-1)/2}dy}{(y^2-1)^{1/2}(1-\tanh^2(v_0)y^{\sqrt{d}})}
\ee

\be
x=\frac{c_1^{1/\sqrt{d}}}{\left( -c_1c_2\right)^{(1+\sqrt{d})/(2\sqrt{d})}\tanh^{(1+\sqrt{d})/\sqrt{d}}(v_0)}\int_1^Y\frac{y^{(\sqrt{d}-3)/2}}{(y^2-1)^{1/2}(1-\tanh^2(v_0)y^{\sqrt{d}})}
\ee

\noindent whereas the action is given by:

\be
S_p=\frac{\left( -c_1c_2\right)^{(1-2\sqrt{d})/(2\sqrt{d})}T}{2\pi\alpha 'c_1^{1/\sqrt{d}}\tanh^{(1+\sqrt{d})/\sqrt{d}}(v_0)}\int_1^Y\frac{y^{(\sqrt{d}+1)/2}dy}{(y^2-1)^{1/2}(1-\tanh^2(v_0)y^{\sqrt{d}})}
\ee

\vspace{0.3cm}

\noindent where $Y=\tanh^{2/\sqrt{d}}(v_{\Lambda})/\tanh^{2/\sqrt{d}}(v_0)$, and $v_0=\sqrt{-c_1c_2}\frac{\sqrt{d}}{2}r_0$, $v_{\Lambda}=\sqrt{-c_1c_2}\frac{\sqrt{d}}{2}\Lambda$.

In Minkowski signature only the integral condition for $L$ is real, while $\mathfrak{R}(\sigma)$ and $S_P$ are complex. We shall work in Euclidean space from now on.
\par
We have not succeeded in solving these integrals in general. When the
sources are located in the vicinity of the singularity, (that is when $\Lambda << l_s$,
these three
integrals can be approximately evaluated by  treating the product $b y$ as
a small quantity, where $b=\tanh(c_1r_0)$. In this case, and working by simplicity in $d=4$, we obtain

\bea
\mathfrak{R}(\sigma)=\frac{b^{1/2}}{\dot t c_1}\Bigg\{\frac{(Y^2-1)^{1/2}(3b^4Y^4+2b^2Y^2-10)}{5Y(1-b^2Y^2)}+\left(\frac{3b^2}{5}+1\right)\int_1^{Y}\frac{z^{1/2}dz}{(z^-1)^{1/2}}\Bigg\}\\
\nonumber\\
\nonumber L=\frac{1}{c_1b^{3/2}}\Bigg\{\frac{3+19b^2-100b^4}{3(40b^4-36b^2+5)}\int_1^Y\frac{dz}{z^{1/2}(z^2-1)^{1/2}}+\hspace{5cm}\\
\nonumber\\
+\frac{(Y^2-1)^{1/2}[(43-100b^4)b^2Y^4+(9-60b^2)Y^3-(10-40b^2)Y^2-(9-60b^2)]}{3Y^{3/2}(1-b^2Y^2)(40b^4-36b^2+5)}\Bigg\}
\eea

\noindent and the potential will be

\bea
\nonumber V=\frac{1}{2\pi\alpha'c_1^2b^{3/2}}\Bigg\{\frac{97b^2-55}{3(36b^2-40b^4+5)}\int_1^Y\frac{dz}{z^{1/2}(z^2-1)^{1/2}}+\hspace{4cm}\\
\nonumber\\
+\frac{(Y^2-1)^{1/2}[(70-109b^2)Y^4-15Y^3-(10-22b^2)Y^2+15]}{Y^{3/2}(1-b^2Y^2)(36b^2-40b^4-5)}\Bigg\}
\eea

\noindent these results are basically the same one obtains from the
 confining background of \cite{ag}, which we have worked out in detail in Appendix A.
\par
A problem with this approximation is that it is not valid when $\Lambda\sim l_s$. We can
improve on it by building the hyperbolic tangent
out of two linear pieces; one corresponding to the confining background,
$a=r$, for $r< l_s$, and another one $a=1$ for $r\geq l_s $

In the region $r> l_s$ the background is Minkowskian (Euclidean), and the action is

\be
S_P=\frac{1}{4\pi\alpha'}\int_Md^2\Sigma\left[ (x')^2+(\dot t)^2+(r')^2\right]
\ee

\noindent  The conserved quantities associated to this system are

\bea
p=x'\\
q=r'\\
j=(r')^2+(x')^2-(\dot t)^2
\eea 

\noindent Now conformal invariance implies

\bea
j=0\\
q=0\\
p=\pm T
\eea

The potential between the two heavy sources can now be computed:

\be
\label{poteuc}
V=\frac{1}{\pi\alpha'}\int_0^Ldx=\frac{1}{\pi\alpha'}L
\ee

\noindent i.e. a linear confining behavior (this is a well-known result;
cf. for example \cite{orlando}).
\par
This behavior can not, however, persist for arbitrarily big values of $\Lambda$
\footnote{ We are grateful to the referee for pointing out this fact},
because
there is another  string configuration which could be energetically preferred
 at great values of $\Lambda$ in which the string goes to the origin and 
comes back to the four dimensional Dirichlet plane. When working directly
 in the approximation in which the hyperbolic tangent in (32) is replaced 
by its asymptotic value (that is equivalent to working in flat space),
 the potential stemming from this configuration is simply

\be
V=\frac{\Lambda}{\pi\alpha'}
\ee

from which it seems clear that when $L$ is much bigger than $\Lambda$ 
it will be energetivally favored. 
\par 
Before computing in a more precise way the potential between 
 heavy sources in this 
configuration, we must check the conditions for this configuration to be
 a solution of the equations of motion in the background defined by 
the metric (\ref{metrlo}). In this case we can work for simplicity 
in the Nambu-Goto framework. The only condition which appears is

\be
x'=0
\ee

Which when inserted in the equations of motion, yields
 a condition for $r(\sigma)$, namely,

\be
r'={\mbox{cotgh}}^{1/2}(c_1r)
\ee

\noindent  giving the measure to be used in the computation of
 the potential,  now reading

\be
V_{scr}=\frac{1}{\pi\alpha'}\int{\mbox{tanh}}^{1/2}(c_1r)=\frac{1}{\pi\alpha'c_1}\left[{\mbox{arctanh}}\left(\sqrt{{\mbox{tanh}}(c_1\Lambda)}\right)-{\mbox{arctan}}\left(\sqrt{{\mbox{tanh}}(c_1\Lambda)}\right)\right]
\ee

We can see a plot of this potential in fig.(\ref{poat}). 

\begin{figure}[h]
\begin{center}
\leavevmode
\epsfxsize=8cm
\epsffile{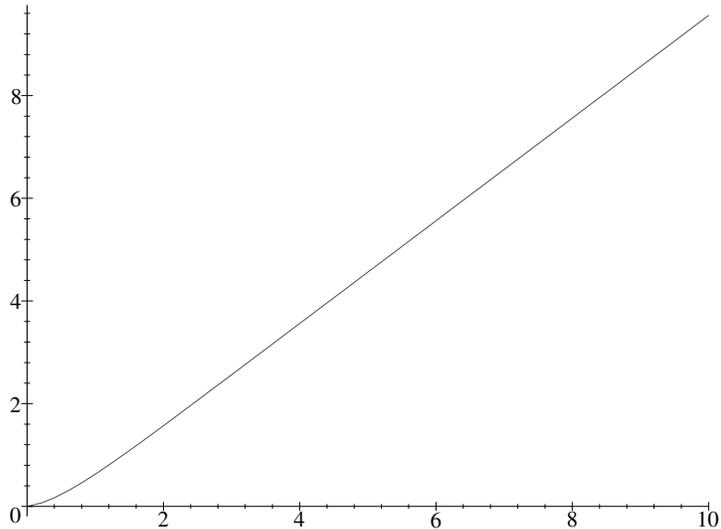}
\caption{\it $\pi\alpha'c_1\, V_{scr}$ vs. $c_1\Lambda$}
\label{poat}
\end{center}  
\end{figure}

This potential can indeed be approximated at great values of $\Lambda$ as 

\be
V_{scr}=\frac{(-0.49+0.99c_1\Lambda)}{\pi\alpha'c_1}\approx\frac{\Lambda}{\pi\alpha'}
\ee

\noindent which is basically the same result worked out previously
 in the Euclidean approximation.

The physical effect of this new configuration is that for any finite value
 of $\Lambda$, there is a transition from the linear confining phase to a 
screening behavior of sorts, characterized by a $L$-independent potential,
the transition taking place at $L\sim \Lambda$.
A similar physical phenomenon appears in the two Wilson loops correlation 
function 
 at finite temperature (\cite{ft}), where beyond certain critical 
distance determined by the size of the loops, and considering the loops 
as circles, the correlation itself vanishes, indicating that the new 
physical configuration consists in a pair of 
disks.
\par
In this linear approximation there is a change in behavior of the loop 
according to the value of $\Lambda$. When $\Lambda> l_s$ the loop does not 
penetrate in the bulk, the world sheet remains flat, and the confining
behavior is minkowskian (this is what we have just found).
\par
There is, however, a  crossover when $\Lambda\sim  l_s$. Then
the loop feels the confining background, penetrates in the bulk, and
actually, a new overconfining $L^3$ branch appears (\cite{ec1}). This 
branch is never
energeticaly favored, however, as  shown in detail in Appendix A.

\, From this point of view, the main difference between the four dimensional 
background presently being studied and the old critical confining one of 
reference \cite{ag} is 
that now all nontrivial 
effects appear when $\Lambda< l_s$ only, and completely dissapear 
for $\Lambda> l_s$, which corresponds almost exactly to the linear dilaton background.

\section{Concluding Remarks}
In this paper we have studied the static potential between heavy sources
in the new four-dimensional background recently introduced in \cite{ag}. 
The dependence on the value of the holographic coordinate at which the 
sources lie (which we have called $\Lambda$) has also been studied, and 
in this particular background there is a crossover at $\Lambda= l_s$ from a 
purely linear behavior towards a region in which a new, cubic branch appears.
This branch has got higher energy cost, however.
\par
As a general comment it could perhaps be said, what has been gained 
with the confining background,
or with the purely four-dimensional one, with respect to the simplest of 
all solutions, namely, the linear dilaton in a flat spacetime?
\par
The answer is twofold: first of all, in order for a holographic interpretation
to be consistent in terms of Callan-Symanzik type of  renormalization
group equations, a mechanism is needed to feed Minkowskian scale 
transformations on the holographic coordinate, and this looks impossible
if the warp factor is trivial (\cite{cs}). In the linear dilaton solution
it is not possible to interpret the behavior of the dilaton in terms
of four dimensional scale transformations $ x^{\m}\rightarrow\lambda x^{\m}$.
\par
The second reason  relies on recent speculations
on the closed tachyon potential \cite{tachyon}. From this point of view
it seems likely that
the confining background is a candidate for the endpoint of tachyon
 condensation starting from the (unstable) linear dilaton.
\par
The candidate renormalization group flow found in reference \cite{agh}
precisely starts from linear dilaton in the region where $ r=\infty$,\footnote{
Which would have been infrared in the confining background, but which here 
corresponds to weak coupling} in $d=4$,
which corresponds to vanishing vacuum expectation value for the tachyon background,
$< T > = 0$. It then flows to the confining background of reference \cite{ag}, but
also in $d=4$, whereas the original solution was critical, i.e., valid in $d=26$ only.
This means that a nonvanishing $< T > \neq 0$ is needed in this region, and this 
is precisely the characteristic of the flow.

\section*{Acknowledgments}
We are insebted to C\'esar G\'omez for useful discussions.
This work ~~has been partially supported by the
European Union TMR program FMRX-CT96-0012 {\sl Integrability,
  Non-perturbative Effects, and Symmetry in Quantum Field Theory} and
by the Spanish grant AEN96-1655.  The work of E.A.~has also been
supported by the European Union TMR program ERBFMRX-CT96-0090 {\sl 
Beyond the Standard model} 
 and  the Spanish grant  AEN96-1664.


\appendix

\section{ Scale Dependence of Wilson Loops in the Confining Background}
We are dealing now with a critical bosonic string moving in a 
background (\cite{ag}) given by

\begin{equation}
ds^2=\phi d\vec x^2+l_c^2d\phi^2
\end{equation}

\noindent where both $x$ and $l_c$ have dimensions of length, and $\phi$ has no dimensions. 

In order to impose the conformal gauge in the world-sheet, we should work in the isothermal coordinate system, that is, if the induced metric takes the form

\begin{equation}
ds^2_{\mbox{ind}}=\left( \phi(x')^2+l_c^2(\phi')^2\right)d\sigma^2+\phi\left(\dot t\right)^2d\tau^2
\end{equation}

\noindent where the prime stands for a derivative with respect to $\sigma$ 
and the dot with respect to $\tau$, and we have defined $t=t(\tau)$, 
$\phi=\phi(\sigma)$, $x=x(\sigma)$ and $ y=z=0$, we look for a coordinate 
transformation of the form

\begin{equation}
\frac{d\sigma}{d\xi}=\dot t\left(\frac{\phi}{\phi(x')^2+l_c^2(\phi')^2}\right)^{1/2}
\end{equation}

\noindent and now, in the variables $(\xi,\tau)$, the metric is proportional to the delta.

The Polyakov action in the coordinate system $(\tau,\sigma)$ reads

\begin{equation}
S_P=\frac{1}{4\pi l_s^2}\int d\sigma d\tau\left[\phi(\dot t)^2+\phi(x')^2+l_c^2(\phi')^2\right]
\end{equation}

\noindent where $\alpha'=l_s^2$ has dimensions of squared length. We can choose $\tau$ and $\sigma$ not to have dimensions, and so we define $t=\tau T$, where now $T$ has dimensions of length.

A conserved quantity  comes from the fact that
 the action does not depend explicitely on $x$, so

\begin{equation}
\label{p}
\phi x'=\mbox{constant}\equiv p
\end{equation}

\noindent otherwise we have the energy-momentum tensor

\begin{equation}
T^\mu_{\hspace{0.2cm}\nu}=\frac{\delta\mathfrak{L}}{\delta\partial_\mu X^\rho}\partial_\nu X^\rho-\eta^\mu_{\hspace{0.2cm}\nu}\mathfrak{L}
\end{equation}

\noindent which now reads

\begin{equation}
T_{01}=0
\end{equation}
\begin{equation}
\label{j}
T_{00}\left(=-T_{11}\right)\equiv j=-\phi T^2+\phi(x')^2+l_c^2(\phi')^2
\end{equation}

\noindent where $p$, in (\ref{p}), has dimensions of length and $j$, in (\ref{j}), of squared length.

We will consider a U-shaped string configuration. Denoting by $\phi_0$ the tip of this U-shaped string we get

\begin{equation}
j=\frac{p^2}{\phi_0}-\phi_0T^2
\end{equation}

\noindent and

\begin{equation}
\label{sigma}
\sigma=\frac{l_c\phi_0^{1/2}}{T}\int_1^Y\frac{y^{1/2}dy}{\sqrt{(y-1)(y+\hat p^2)}}
\end{equation}

\noindent this can be written for the coordinate $\xi$ as

\begin{equation} 
\label{xi} 
\xi=\frac{l_c\phi_0^{1/2}}{T}\int_1^Y\frac{(y+\hat j)^{1/2}dy}{\sqrt{(y-1)(y+\hat p^2)}}
\end{equation}

\noindent where we have defined $j=\hat j\phi_0T^2$ and $p^2=\hat p^2\phi_0^2T^2$ so they follow the relation

\begin{equation}
\hat j=\hat p^2-1
\end{equation}

Now we can use (\ref{sigma}) to get from (\ref{p})

\begin{equation}
\label{x}
x=l_c\hat p\phi_0^{1/2}\int_1^Y\frac{dy}{\sqrt{y(y-1)(y+\hat p^2)}}
\end{equation}

Imposing Dirichlet boundary conditions at the codimension one hypersurface $\phi=\Lambda$, so we get from (\ref{sigma}) and (\ref{x})

\begin{equation} 
\label{sig} 
\mathfrak{R}(\xi)=\frac{2l_c\phi_0^{1/2}}{T}\int_1^{\sqrt{\Lambda/\phi_0}}\sqrt{\frac{\chi^2(\chi^2+\hat j)}{(\chi^2-1)(\chi^2+\hat p^2)}}d\chi
\end{equation}

\noindent where $\mathfrak{R}(\xi)$ is defined through 
$0\leq\xi\leq\mathfrak{R}(\xi)$, and the physical separation of the 
heavy sources (which is the only external data) must be now be fed through:
\be 
L=\int_{0}^{\mathfrak{R}(\xi)}\frac{dx}{d\xi}d\xi
\ee
that is,
\begin{equation}
\label{l}
L=2l_c\hat p\phi_0^{1/2}\int_1^{\sqrt{\Lambda/\phi_0}}\frac{d\chi}{\sqrt{(\chi^2-1)(\chi^2+\hat p^2)}}
\end{equation}

The action, in isothermal coordinates, is then given by

\begin{equation} 
\label{s} 
S=\frac{\hat jl_ct\phi_0^{3/2}}{2\pi l_s^2}\int_1^{\sqrt{\Lambda/\phi_0}}\sqrt{\frac{\chi^2(\chi^2+\hat j)(2\chi^2+\hat j)^2}{(\chi^2-1)(\chi^2+\hat p^2)}}d\chi 
\end{equation} 

Before trying to solve these equations we have got one extra condition for conformal invariance, that is the vanishing of the energy-momentum tensor. This condition can be seen as

\begin{eqnarray}
\label{con}
\nonumber \hat j=0\\
\hat p^2=1
\end{eqnarray}

\noindent with these two conditions one can write (\ref{sig}), (\ref{l}) and (\ref{s}) as

\begin{eqnarray}
\mathfrak{R}(\xi)=\mathfrak{R}(\sigma)=\frac{2l_c\phi_0^{1/2}}{T}\int_1^{\sqrt{\Lambda/\phi_0}}\frac{\chi^2d\chi}{\sqrt{\chi^4-1}}\\
L=\pm 2l_c\phi_0^{1/2}\int_1^{\sqrt{\Lambda/\phi_0}}\frac{d\chi}{\sqrt{\chi^4-1}}\\
S=\frac{l_ct\phi_0^{3/2}}{\pi l_s^2}\int_1^{\sqrt{\Lambda/\phi_0}}\frac{\chi^4d\chi}{\sqrt{\chi^4-1}}
\end{eqnarray}

These integrals can be solved in terms of elliptic functions as

\begin{eqnarray}
\label{cond}
\mathfrak{R}(\sigma)=\frac{2l_c\phi_0^{1/2}}{T}\left[ \frac{\sqrt{2}}{2}F\left( \cos^{-1}\sqrt{\frac{\phi_0}{\Lambda}},\frac{1}{\sqrt{2}}\right)-\sqrt{2}E\left( \cos^{-1}\sqrt{\frac{\phi_0}{\Lambda}},\frac{1}{\sqrt{2}}\right)+\sqrt{\frac{\phi_0}{\Lambda}}\sqrt{\frac{\Lambda^2}{\phi_0^2}-1}\right]\\
\nonumber\\
\label{cond2}
& \hspace{-15cm} L=\pm 2l_c\phi_0^{1/2}\frac{1}{\sqrt{2}}F\left( \cos^{-1}\sqrt{\frac{\phi_0}{\Lambda}},\frac{1}{\sqrt{2}}\right)\\
\nonumber\\
\label{cond3}
& \hspace{-15cm} S=\frac{l_cT\phi_0^{3/2}}{\pi l_s^2}\left[ \frac{1}{3\sqrt{2}}F\left( \cos^{-1}\sqrt{\frac{\phi_0}{\Lambda}},\frac{1}{\sqrt{2}}\right)+\frac{1}{3}\sqrt{\frac{\phi_0}{\Lambda}}\sqrt{\frac{\Lambda^2}{\phi_0^2}-1}\right]
\end{eqnarray}

This leads to essentially the same equations than
 those in \cite{ec1}, taking into account the $1/4\pi$ factors in the 
definition of the Polyakov action and the $1/2\pi$ in the Nambu-Goto one, 
but we also obtain another extra condition, namely, that of (\ref{cond}). 
Together, these three equations imply that

\begin{eqnarray} 
\label{acc} 
S=\frac{TE\left( \cos^{-1}\sqrt{\frac{\phi_0}{\Lambda}},\frac{1}{\sqrt{2}}\right)}{6\pi l_s^2l_c^2F^3\left( \cos^{-1}\sqrt{\frac{\phi_0}{\Lambda}},\frac{1}{\sqrt{2}}\right)}L^3+\\
& \hspace{-5cm}\nonumber+\frac{TL^2}{12\pi l_s^2l_c^2F^2\left( \cos^{-1}\sqrt{\frac{\phi_0}{\Lambda}},\frac{1}{\sqrt{2}}\right)}\left( T\mathfrak{R}(\sigma)-2l_c\Lambda^{1/2}\sqrt{1-\frac{\phi_0^2}{\Lambda^2}}\right)+\frac{Tl_c\Lambda^{3/2}}{3\pi l_s^2}\sqrt{1-\frac{\phi_0^2}{\Lambda^2}}
\end{eqnarray}

\begin{figure}[h]
\begin{center}
\leavevmode
\epsfxsize=8cm
\epsffile{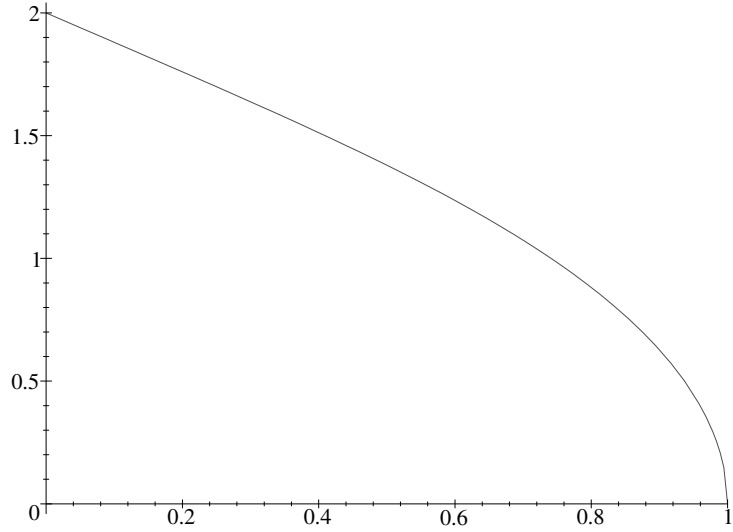}
\caption{\it $T\mathfrak{R}(\sigma)/l_c\sqrt{\Lambda}$ vs. $\sqrt{\phi_0/\Lambda}$ }
\label{f3}
\end{center}  
\end{figure}

Before trying to extract any conclusion from this expression, we may see what (\ref{cond}) is telling. 

From this equation we can read a relation between $T\mathfrak{R}(\sigma)/l_c\sqrt{\Lambda}$ and $\sqrt{\phi_0/\Lambda}$, plotted in fig.(\ref{f3}). This relation sets the range of $\sigma$ as 

\begin{equation}
0\leq\frac{\mathfrak{R}(\sigma)}{\Lambda^{1/2}}\leq\frac{2l_c}{T}  
\end{equation}

\noindent so, in the limit $\Lambda\rightarrow\infty$, this variable diverges. This divergence is also present in (\ref{acc}), but is exactly cancelled, so in this limit we find the following relation for the  potential between heavy sources:

 \begin{equation}
  \label{eq:acin}
  V=\frac{E(1/\sqrt{2})}{3\pi l_s^2l_c^2K(1/\sqrt{2})^3}L^3+\frac{2l_c\Lambda^{3/2}}{3\pi l_s^2}
\end{equation}

So we find an over confinig region and two different divergences, but only one of them is present in the potential.

Now we can compute the potential in the limit $\Lambda\approx\phi_0$. In a na\"\i ve way one obtains

\begin{equation}
V=\frac{L\Lambda}{3\pi l_s^2}
\end{equation}

\noindent i.e. a linear confining behaviour.

There is another interesting relation, that of (\ref{cond2}). This
equation give us the relation between $\sqrt{\frac{\phi_0}{\Lambda}}$
and $\frac{L}{l_c\sqrt{\Lambda}}$. This is plotted is fig.(\ref{f2})
and tell us that for each $L$, there are two possible configurations,
which we will call regions $I$ and $II$ (\cite{ec1}\cite{kogan}).

\begin{figure}[h]
\begin{center}
\leavevmode
\epsfxsize=12cm
\epsffile{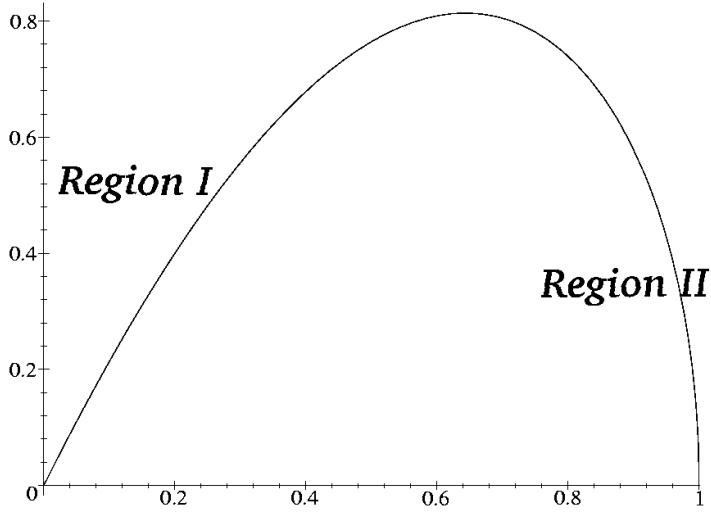}
\caption{\it Relation between $x=\sqrt{\frac{\phi_0}{\Lambda}}$ and $y=\frac{L}{l_c\sqrt{\Lambda}}$}
\label{f2}
\end{center}
\end{figure}

In order to guess which configuration is in, favored energetically we have to look for analytic (up to an order of magnitude) expressions of the equations (\ref{cond}-\ref{cond3}). Firstly we expand the elliptic functions as

\begin{eqnarray}
  \label{F}
\nonumber F(\cos^{-1}(x),1/\sqrt{2})=\frac{2}{\pi}K(1/\sqrt{2})\left[ \cos^{-1}(x)-x\sqrt{1-x^2}\left(\frac{5}{3}-\frac{2}{3}x^2\right)\right]+\hspace{0.5cm}\\
+x\sqrt{1-x^2}\left(\frac{7}{4}-\frac{3}{4}x^2\right)
\end{eqnarray}
\begin{eqnarray}
  \label{E}
\nonumber E(\cos^{-1}(x),1/\sqrt{2})=\frac{2}{\pi}E(1/\sqrt{2})\left[ \cos^{-1}(x)-x\sqrt{1-x^2}\left(\frac{5}{3}-\frac{2}{3}x^2\right)\right]+\hspace{0.5cm}\\
+x\sqrt{1-x^2}\left(\frac{19}{12}-\frac{7}{12}x^2\right)
\end{eqnarray}

\noindent with these two equations we can invert (\ref{cond2}). We obtain four solutions to the equation, but only two of them are physically relevant. These are

\begin{eqnarray}
\label{sol1}
\left(\frac{\phi_0^{(1)}}{\Lambda}\right)^{1/2}=\frac{5}{4}+\frac{1}{12}\sqrt{B}-\frac{1}{12}\sqrt{6}\sqrt{\frac{43A^{1/3}\sqrt{B}-A^{2/3}\sqrt{B}+36\sqrt{B}z^2-64\sqrt{B}+405A^{1/3}}{A^{1/3}\sqrt{B}}}
\end{eqnarray}
\begin{eqnarray}
\label{sol2}
\left(\frac{\phi_0^{(2)}}{\Lambda}\right)^{1/2}=\frac{5}{4}-\frac{1}{12}\sqrt{B}+\frac{1}{12}\sqrt{6}\sqrt{\frac{43A^{1/3}\sqrt{B}-A^{2/3}\sqrt{B}+36\sqrt{B}z^2-64\sqrt{B}+405A^{1/3}}{A^{1/3}\sqrt{B}}}
\end{eqnarray}

\noindent where $z=\frac{L}{l_c\Lambda^{1/2}}$, and

\begin{eqnarray}
\label{a}
A=-1161z^2+512+9\sqrt{576z^6+13569z^4-9216z^2}
\end{eqnarray}
\begin{eqnarray}
\label{b}
B=-\frac{-129A^{1/3}-6A^{2/3}+216z^2-384}{A^{1/3}}
\end{eqnarray}

\noindent each solution corresponds to one of the two regions in fig.(\ref{f2}), namely, (\ref{sol1}) corresponds to \textsl{region I}, while (\ref{sol2}) corresponds to \textsl{region II}.

An important feature of these solutions is that they are valid in all the range of $\phi_0\leq\phi\leq\Lambda$. Given this, we can write down an expression for the potential valid in all the range of $\phi_0\leq\phi\leq\Lambda$ for each region. This is given by

\begin{eqnarray}
\frac{3\pi l_s^2}{2l_c\Lambda^{3/2}}V^{(1)}=\sqrt{1-\left(\frac{5}{4}+\frac{\sqrt{B}}{12}-\frac{C_+}{12}\right)^4}+\frac{\sqrt{3}}{3}\left(\frac{5}{4}+\frac{\sqrt{B}}{12}-\frac{C_+}{12}\right)^3\times\\
\times\sqrt{\left(\frac{5}{4}+\frac{\sqrt{B}}{12}-\frac{C_+}{12}\right)^2-\frac{9}{4}-\frac{5\sqrt{B}}{12}+\frac{5C_+}{12}}
\end{eqnarray}

\noindent for region I, and

\begin{eqnarray}
\frac{3\pi l_s^2}{2l_c\Lambda^{3/2}}V^{(2)}=\sqrt{1-\left(\frac{5}{4}-\frac{\sqrt{B}}{12}+\frac{C_-}{12}\right)^4}+\frac{\sqrt{3}}{3}\left(\frac{5}{4}-\frac{\sqrt{B}}{12}+\frac{C_-}{12}\right)^3\times\\
\times\sqrt{\left(\frac{5}{4}-\frac{\sqrt{B}}{12}+\frac{C_-}{12}\right)^2-\frac{9}{4}+\frac{5\sqrt{B}}{12}-\frac{5C_-}{12}}
\end{eqnarray}

\noindent for region II. Where $A$ and $B$ are given by (\ref{a}) and (\ref{b}) respectively and

\begin{equation}
C_\pm=\sqrt{6}\sqrt{\frac{43A^{1/3}\sqrt{B}-A^{2/3}\sqrt{B}+36\sqrt{B}z^2-64\sqrt{B}\pm 405A^{1/3}}{A^{1/3}\sqrt{B}}}
\end{equation}

These solutions are plotted in fig.(\ref{f6}). From this plot we can see that to region II corresponds a smaller potential energy.

\begin{figure}[h]
\begin{center}
\leavevmode
\epsfxsize=8cm
\epsffile{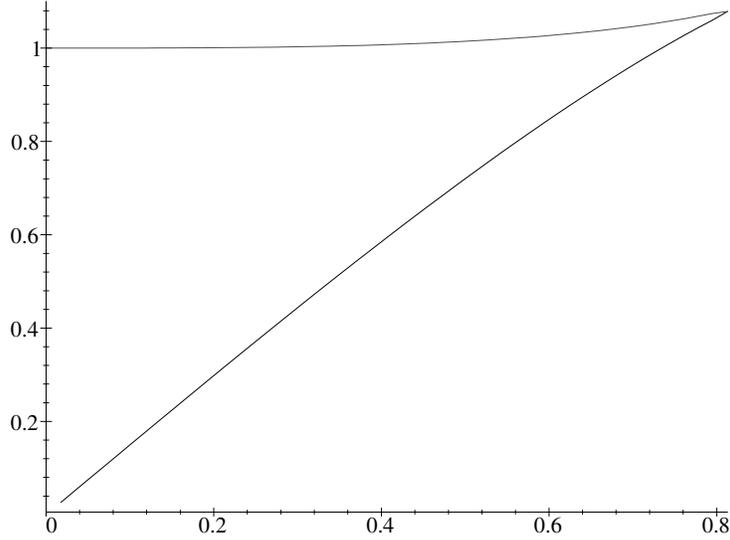}
\caption{\it $j\bar j$ potential corresponding to the two available configurations. The plot is $\frac{3\pi l_s^2}{2l_c\Lambda^{3/2}}V$ {\sl{vs.}} $\frac{L}{l_c\Lambda^{1/2}}$}
\label{f6}
\end{center}
\end{figure}

An interesting feature of this plot is that the potential corresponding to the linear confining region does not reach the point $L=0$. This can be interpreted as follows.

Looking at fig.(\ref{f2}), it is apparent that there are two critical points in the holographic direction: $\phi_0=0$ 
and $\phi_0=\infty$, however, in view of fig.(\ref{f2}), we can see that 
there is another one, the corresponding to the maximun in that figure. 
This point corresponds to the maximun  potential energy 
available between the two sources, that is, to the point of 
maximun possible separation between the two sources.

This presumably means that this is the point where a new pair $j\bar j$ is created.

The two regions in fig.(\ref{f2}) correspond to both sides of the critical plane just defined. The distance between the sources decreases as we move away from this plane.

In this way, as we approach $\phi_0=0$, we move towards $L=0$. However, in this limit, the metric is singular, so we can never reach $L=0$ in this way. In fact the string, through $\phi_0$, would go into the singularity before we reach it, through $\Lambda$.

This is what makes this configuration of region $II$ disappear, leaving us with just region $I$, as $\Lambda\rightarrow\infty$. 
\par
There is however in this limit another point that must be clarified.

In the body of the paper, we have demonstrated that the equations of
motion that arise from Nambu-Goto and Polyakov actions are the same,
being this last written in the isothermal coordinate system. However,
in the Polyakov's case we found an extra condinition, namely, that of
$\sigma$. This condition plays a role more important than been merely
a parametrization of the world-sheet.

The divergence in $\sigma$, as seen from eq.(\ref{cond}), makes the
finite part of the potential in eq.(\ref{eq:acin}) to be different
from the one computed in \cite{ec1}. This fact does not point to a 
contradiction, however. One
must simply have in mind that the equivalence of the equations of
motion coming from different lagrangians is due to the fact that one
can always add a total derivative of a function of position and time
to the old lagrangian.

When saying this one always assumes that such a function is well behaved
 at the boundaries. Now this is not the case, because in
the boundaries is where we are in trouble: a geometric singularity at the
origin  and a divergent
 coupling constant at the infinity of the holographic coordinate.

\section{ Detailed Analysis of some Supergravity Solutions}
It is intesting to consider related backgrounds which, although not
conformal, appear naturally in the context of supergravity. Let us choose
as an example the one in reference \cite{porrati}.
This background is given by

\be
ds^2=y^{1/2}d\vec{x}^2 +Ady^2
\ee

The induced metric will be

\be
ds^2_{\mbox{ind}}=\left[ y^{1/2}(x')^2+A(y')^2\right] d\sigma^2+y^{1/2}(\dot t)^2d\tau^2
\ee

\noindent so the Polyakov action is

\be
\label{suac}
S_P=\frac{1}{4\pi\alpha'}\int_Md^2\Sigma\Big\{ y^{1/2}\left[ (x')^2+(\dot t)^2\right]+A(y')^2\Big\}
\ee

The change of variables that should be done in order to compute in the isothermal coordinate system will be

\be
\frac{d\sigma}{d\xi}=T\frac{y^{1/4}}{\left[ j+y^{1/2}T^2\right]^{1/2}}
\ee

\noindent where $j$ is again the ''$00$'' component of the energy-momentum tensor obtained from the action (\ref{suac})

\be
j=y^{1/2}\left[ (x')^2-(\dot t)^2\right]+A(y')^2
\ee

\noindent the other conserved quantitie associated to (\ref{suac}) is

\be
p=y^{1/2}x'
\ee

\noindent from these two quantities we can define a new set of constants

\bea
j=\hat jy_0^{1/2}T^2\\
p^2=\hat p^2y_0T^2
\eea

\noindent where $y_0$ is defined, as is customary, as the tip of the U-shaped string configuration. Conformal invariance implies

\bea
\hat j=0\\
\hat p^2=1
\eea

Now we are ready to construct the constraint conditions to the parametrization of the world-sheet and to the lenght of the wilson loop as

\bea
\mathfrak{R}(\xi)=\frac{4A^{1/2}y_0^{3/4}}{T}\int_1^{(\Lambda/y_0)^{1/4}}\frac{\chi^4d\chi}{\left[ (\chi^2-1)(\chi^2+1)\right]^{1/2}}\\
L=4A^{1/2}y_0^{3/4}\int_1^{(\Lambda/y_0)^{1/4}}\frac{\chi^2d\chi}{\left[ (\chi^2-1)(\chi^2+1)\right]^{1/2}}
\eea

\noindent and the action is

\be
S_P=\frac{2TA^{1/2}y_0^{5/4}}{\pi}\int_1^{(\Lambda/y_0)^{1/4}}\frac{\chi^6d\chi}{\left[ (\chi^2-1)(\chi^2+1)\right]^{1/2}}
\ee

\begin{figure}[h]
\begin{center}
\leavevmode
\epsfxsize=8cm
\epsffile{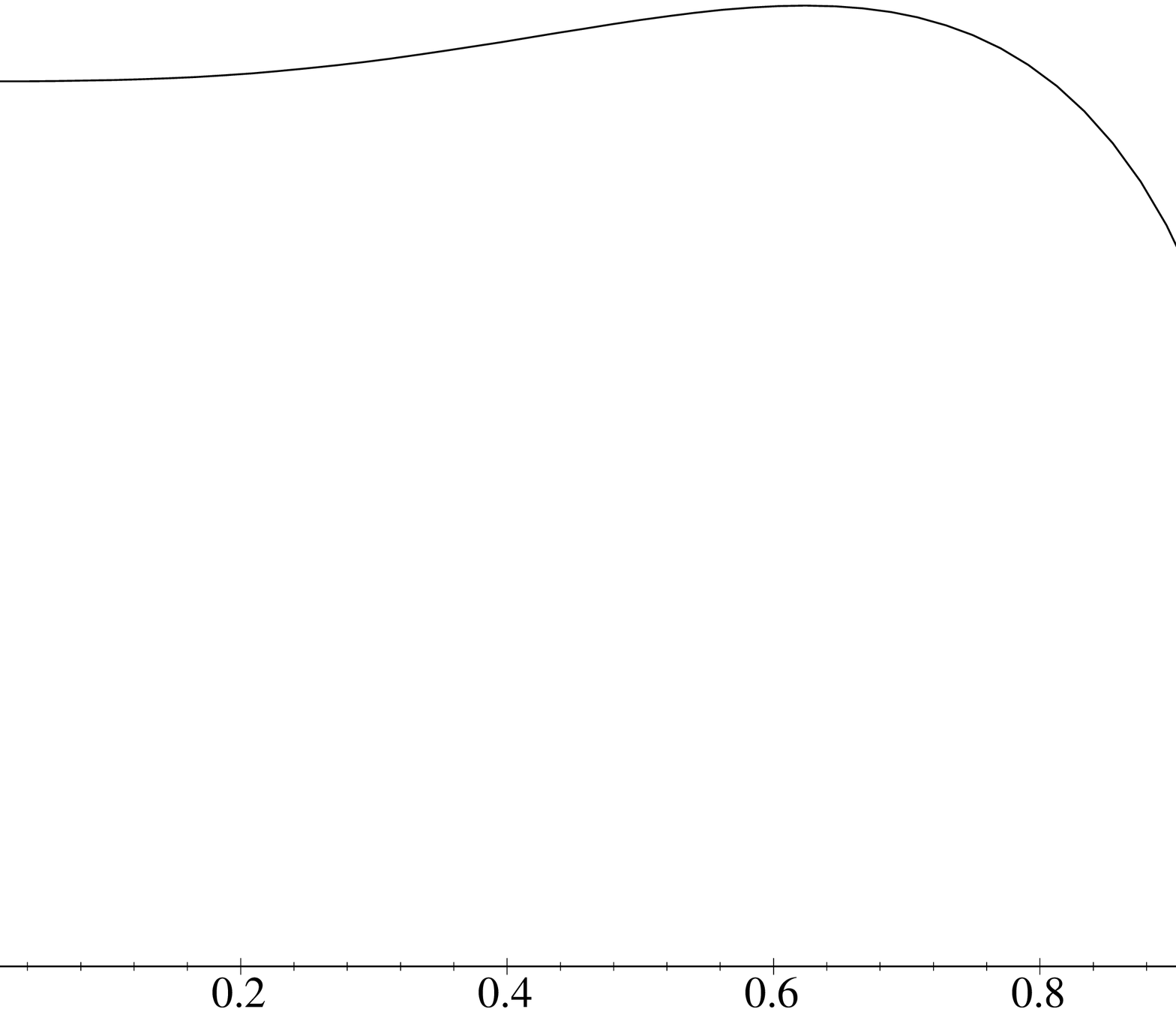}
\caption{\it $\frac{3T\mathfrak{R}(\sigma)}{4A^{1/2}\Lambda^{3/4}}$ {\sl{vs.}} $\left(\frac{y_0}{\Lambda}\right)^{1/4}$}
\label{suf2}
\end{center}
\end{figure}

These three integrals can be solved as

\be
\label{susig}
\mathfrak{R}(\xi)=\frac{4A^{1/2}y_0^{3/4}}{3T}\left\{\frac{1}{\sqrt{2}}F\left( \arccos\left(\frac{y_0}{\Lambda}\right)^{1/2},\frac{1}{\sqrt{2}} \right)+\frac{\Lambda^{3/4}}{y_0^{3/4}}\sqrt{1-\frac{y_0}{\Lambda}}\right\}
\ee
\be
\label{sul}
L=4A^{1/2}y_0^{3/4}\left\{\frac{1}{\sqrt{2}}F\left( \arccos\left(\frac{y_0}{\Lambda}\right)^{1/2},\frac{1}{\sqrt{2}} \right)-\sqrt{2}E\left( \arccos\left(\frac{y_0}{\Lambda}\right)^{1/2},\frac{1}{\sqrt{2}} \right)+\frac{\Lambda^{1/4}}{y_0^{1/4}}\sqrt{1-\frac{y_0}{\Lambda}}\right\}
\ee
\bea
\label{supot}
\nonumber S_P=\frac{2TA^{1/2}y_0^{5/4}}{5\pi l_s^2}\Biggr\{\frac{3}{\sqrt{2}}F\left( \arccos\left(\frac{y_0}{\Lambda}\right)^{1/2},\frac{1}{\sqrt{2}} \right)-3\sqrt{2}E\left( \arccos\left(\frac{y_0}{\Lambda}\right)^{1/2},\frac{1}{\sqrt{2}} \right)+\\
+3\frac{\Lambda^{1/4}}{y_0^{1/4}}\sqrt{1-\frac{y_0}{\Lambda}}+\frac{\Lambda^{5/4}}{y_0^{5/4}}\sqrt{1-\frac{y_0}{\Lambda}}\Biggr\}
\eea

In the limit $\Lambda\approx y_0$ we can find the following behaviour

\be
V=\frac{3L\Lambda^{1/2}}{5\pi l_s^2}
\ee

\begin{figure}[h]
\begin{center}
\leavevmode
\epsfxsize=8cm
\epsffile{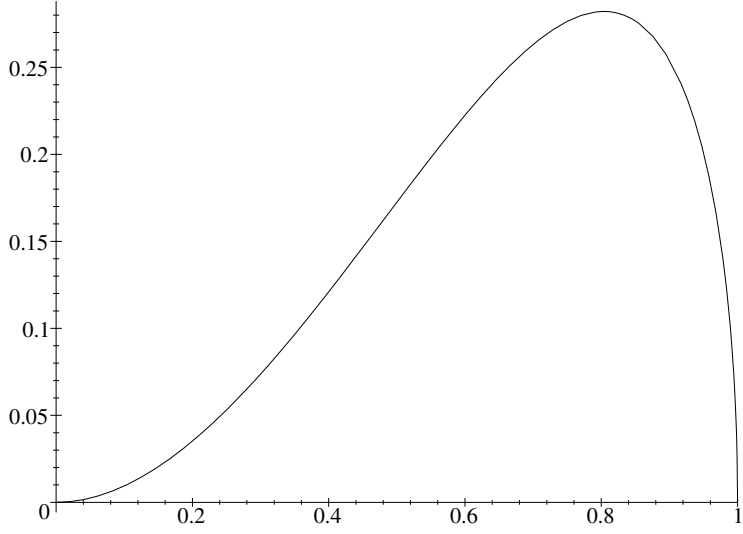}
\caption{\it $\frac{L}{4A^{1/2}\Lambda^{3/4}}$ {\sl{vs.}} $\left(\frac{y_0}{\Lambda}\right)^{1/4}$}
\label{suf1}
\end{center}
\end{figure}

The first thing one may say about this background is that it, basically, give us the same kind of configurations seen in the last section. First of all, the range of $\xi$ is read from fig.(\ref{suf2}). This is

\be
0\leq\frac{\mathfrak{R}(\xi)}{\Lambda^{3/4}}\leq\frac{4A^{1/2}}{3T}
\ee

\noindent This behaviour of $\sigma$ allow us to write the potential when $\Lambda\rightarrow\infty$ as

\be
V=\frac{4A^{1/2}}{5\pi\l_s^2}\Lambda^{5/4}
\ee

Secondly, the string can live in two possible configurations as seen in fig.(\ref{suf1})

Now we want to say which configuration costs less energy. In order to do so, we should invert eq.(\ref{sul}), as in the previous section. However, in this case, the more we can do is an interpolation to get a polynomial solution. Once again, the region corresponding to a linear confining region is the physically relevant. This turns out to be

\be
\frac{5\pi l_s^2}{4A^{1/2}\Lambda^{5/4}}V=(1+3\Delta^4)\sqrt{1-\Delta^4}-3\Delta^5\sqrt{\Delta-\Delta^2}
\ee

\begin{figure}[h]
\begin{center}
\leavevmode
\epsfxsize=8cm
\epsffile{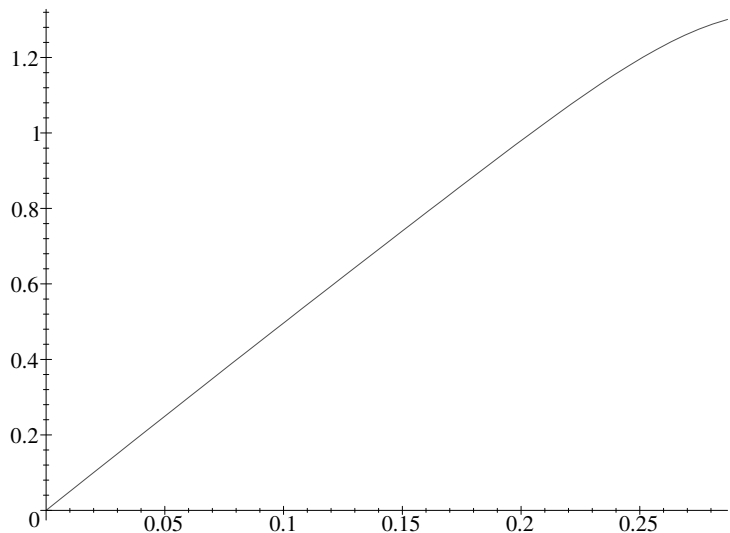}
\caption{\it $\frac{5\pi l_s^2}{4A^{1/2}\Lambda^{5/4}}V$ {\sl{vs.}} $\frac{L}{4A^{1/2}\Lambda^{1/4}}$}
\label{suf3}
\end{center}
\end{figure}

\noindent where $\Delta$ is given, approximately, by

\be
\Delta=-129.22z^6+19.18z^5-5.33z^4+0.42z^3-z^2+0.28\times 10^{-5}z+1
\ee

\noindent where $z=\frac{L}{4A^{1/2}\Lambda^{1/4}}$. 

This potential is given in fig.(\ref{suf3}). There are a few comments
to be made. First of all we see that the potential is linear in the
product $L\Lambda^{1/2}$, corresponding to the so called region $II$.
However, in this case we cannot see the effect of the singularity in
the origin, due to the interpolation we made, altough it is expected to
cause similar problems as with the confining background.


\end{document}